\documentclass[10pt,a4j]{article}
\usepackage[dvips]{graphicx}
\usepackage{mathrsfs}
\usepackage{amsthm} 
\usepackage{amsmath}
\usepackage{amssymb}
\usepackage{amsfonts}
\usepackage{wrapfig}
\usepackage{cite}
\usepackage{chemarrow}

\begin{document}

\renewcommand{\figurename}{\small{Fig.}~}
\renewcommand{\labelitemi}{}
\renewcommand{\thefootnote}{$\dagger$\arabic{footnote}}
\renewcommand{\footnoterule}{%
  \vspace{2pt}                      
 \noindent\rule{6.154cm}{0.4pt}   
  \vspace{4pt}                     
}
\pagestyle{plain}

\begin{flushright}
\textit{Entropy Change in the Adiabatic Process}
\end{flushright}
\vspace{2mm}

\begin{center}
\setlength{\baselineskip}{25pt}{\LARGE\textbf{Entropy Change in the Carnot Cycle}}
\end{center}
\vspace{-6mm}
\begin{center}
\setlength{\baselineskip}{25pt}{\large\textbf{Entropy in the Adiabatic Process}}
\end{center}
\vspace*{4mm}
\begin{center}
\large{Kazumi Suematsu} \vspace*{2mm}\\
\normalsize{\setlength{\baselineskip}{12pt} 
Institute of Mathematical Science\\
Ohkadai 2-31-9, Yokkaichi, Mie 512-1216, JAPAN\\
E-Mail: suematsu@m3.cty-net.ne.jp,  Tel/Fax: +81 (0) 593 26 8052}\\[8mm]
\end{center}@

\hrule
\vspace{0mm}
\begin{flushleft}
\textbf{\large Abstract}
\end{flushleft}
Entropy change in the Carnot cycle is discussed. In particular, the isentropic change in the adiabatic expansion (or compression) is reinvestigated.
\begin{flushleft}
\textbf{\textbf{Key Words}}: Carnot Cycle/ Adiabatic Process/ Entropy Change/
\end{flushleft}
\hrule
\vspace{3mm}
\setlength{\baselineskip}{13pt}
\section{Introduction}

\begin{wrapfigure}[18]{r}{7cm}
\vspace*{-0.5mm}
\begin{center}
\includegraphics[width=6.5cm]{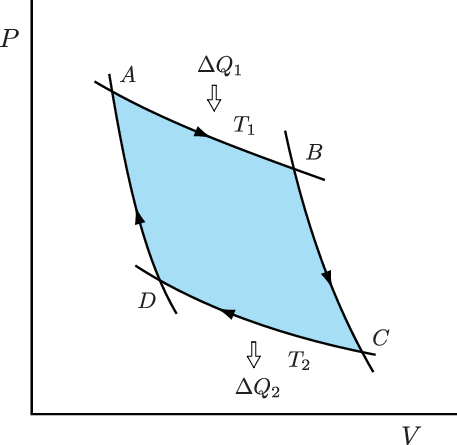}
\end{center}
\vspace*{2mm}
\setlength{\baselineskip}{10.5pt}{\small  Fig. 1: Carnot Cycle on the PV Diagram.}\label{CarnotCycle}
\end{wrapfigure}

Consider a heat engine filled with ideal gas molecules. Let the engine follow the Carnot cycle which is constructed from the known four processes: 
\begin{eqnarray}
A(P_{1}, V_{1}, T_{1})\hspace{5mm} \autorightarrow{Isothermal}{$\Delta Q_{1}$} \hspace{5mm} B(P_{2}, V_{2}, T_{1}) \hspace{5mm} \autorightarrow{Adiabatic}{$\Delta Q=0$} \hspace{5mm}  C(P_{3}, V_{3}, T_{2})\notag\\[4mm]
\autorightarrow{Isothermal}{$\Delta Q_{2}$} \hspace{5mm} D(P_{4}, V_{4}, T_{2}) \hspace{5mm} \autorightarrow{Adiabatic}{$\Delta Q=0$} \hspace{5mm}  A(P_{1}, V_{1}, T_{1})\label{1-1}
\end{eqnarray}
(see Fig. \ref{CarnotCycle}). It is assumed that these processes are carried out under the reversible (quasi-static) conditions. Then the entropy $S$ is defined by the differential form $dS=dQ/T$. So the entropy was originally introduced on the basis of the premise of the reversible (quasi-static) change.

It has been well established that $S$ is a state function, and for this reason, $\Delta S$ is invariable, whether a path that links given two states is reversible or irreversible, if the initial and the final states are in equilibrium. Hence, for any cyclic process we must have
\begin{equation}
\oint dS=0\label{1-2}
\end{equation}
In particular, for the Carnot cycle under discussion, we have
\begin{equation}
\oint dS=\frac{\Delta Q_{1}}{T_{1}}+\frac{\Delta Q_{2}}{T_{2}}=0\label{1-3}
\end{equation}
The question is the first equality:
\begin{equation}
\oint dS=\frac{\Delta Q_{1}}{T_{1}}+\frac{\Delta Q_{2}}{T_{2}}\label{1-4}
\end{equation}
The equality (\ref{1-4}) is based on the general acceptance that $\Delta S_{B\rightarrow C}=\Delta S_{D\rightarrow A}=0$, which is justified by the fact that no heat exchange occurs in the adiabatic process. This is logically true; in this sense there is no room to doubt eq. (\ref{1-4}). However we note that the entropy is a state function and there is a change of state in the adiabatic process \rule[0.5mm]{0.7cm}{0.5pt} for instance, such that $B(P_{2}, V_{2}, T_{1})\rightarrow C(P_{3}, V_{3}, T_{2})\, $\rule[0.5mm]{0.7cm}{0.5pt}, so that it seems natural to consider that some change of the entropy may occur during the process, and hence to require a more rigorous proof for $\Delta S_{adiabatic}=0$.  Unfortunately, no inquiry into this problem has so far been made in the text books\cite{Clausius, Guggenheim, Planck, Sommerfeld, Glasstone, Moor, Kittel, Denbigh, Callen, Reif, Gibbs, Atkins, Prigogine, Schroeder}. This is the reason the author submitted this short article.
 
Thermodynamics is a deep physics. Through the analysis of the Carnot cycle, an idealized model of the steam engine, the amazing theorem in theoretical physics, the second law ($\Delta S\ge 0$), was deduced, having laid the foundation of chemical thermodynamics. Because of the depth and the vastness, for its full comprehension we must study repeatedly the series of basics from the classical to the statistical thermodynamics. In such circumstances, it seems essential to discuss more rigorously the validity of eq. (\ref{1-4}).

\section{Entropy Change in Adiabatic Process}
To examine whether the change of the entropy occurs in the adiabatic process, let us consider the change of state of  $B(P_{2}, V_{2}, T_{1})\rightarrow C(P_{3}, V_{3}, T_{2})$. Since $S$ is a state function, the path $B\rightarrow C$ is separable, and we may construct an alternative route to the state $C$, for instance
\begin{eqnarray}
B(P_{2}, V_{2}, T_{1})\hspace{5mm} \autorightarrow{$\Delta Q=\Delta W=0$}{$V_{2}\rightarrow V_{3}$} \hspace{5mm} X(P_{X}, V_{3}, T_{1}) \hspace{5mm} \autorightarrow{$\Delta V=0$}{$T_{1}\rightarrow T_{2}$} \hspace{5mm}  C(P_{3}, V_{3}, T_{2})\label{1-5}
\end{eqnarray}

\noindent (a) Let the path $B\rightarrow X$ be an expansion process with no work and no heat exchange. The free expansion of gas molecules into the vacuum will correspond to this case. In that case the volume change occurs, but $\Delta Q=0$, $\Delta W=0$ and $\Delta U=0$ ($P_{X}$, of course, satisfies the equality: $P_{X}=(V_{2}/V_{3})P_{2}$).\\
(b) Let the path $X\rightarrow C$ be a process, during which the system releases $\Delta Q$, but has no volume change.\\[-1mm]

For the path $B\rightarrow X$, according to the Boltzmann formula, there is the entropy change of the form:
\begin{equation}
\Delta S_{B\rightarrow X}=nR\log \frac{V_{3}}{V_{2}}\label{1-6}
\end{equation}
while for the path $X\rightarrow C$, there is the temperature drop $T_{1}\rightarrow T_{2}$ of the system through the release of $\Delta Q$ to the heat reserver, which leads to the decrease of entropy.\\[-1mm]

Since $U$ is a state function, we have
\begin{equation}
\Delta U_{B\rightarrow X\rightarrow C}= \Delta U_{B\rightarrow C}\notag
\end{equation}
Comparing the two routes, $B\rightarrow X\rightarrow C$ and $B\rightarrow C$ , it must be that 
\begin{equation}
\Delta U=\left(\Delta Q\right)_{X\rightarrow C}=\left(\int-P dV\right)_{B\rightarrow C} \hspace{5mm}(\textit{equivalent heat})\label{1-7}
\end{equation}
We may express eq. (\ref{1-7}) in the differential form: $dQ=-PdV$ (\textit{equivalent heat}). And we have
\begin{equation}
\Delta S_{X\rightarrow C}=\int_{T_{1}}^{T_{2}}\frac{dQ}{T}=\int_{V_{2}}^{V_{3}} -\frac{P}{T}dV\label{1-8}
\end{equation}
The total entropy change throughout the adiabatic process thus becomes
\begin{equation}
\Delta S_{B\rightarrow X\rightarrow C}=\Delta S_{B\rightarrow X}+\Delta S_{X\rightarrow C}=nR\log \frac{V_{3}}{V_{2}}-\int_{V_{2}}^{V_{3}}\frac{P}{T}dV\label{1-10}
\end{equation}
Applying the equation of state, $PV=nRT$, we arrive at the known result for the adiabatic process:
\begin{equation}
\Delta S_{B\rightarrow X\rightarrow C}\equiv\Delta S_{B\rightarrow C}=nR\log \frac{V_{3}}{V_{2}}-nR\log \frac{V_{3}}{V_{2}}=0\label{1-11}
\end{equation}
As we can see, the increase of the entropy due to the volume expansion during the adiabatic process is exactly canceled out by the decrease of the entropy due to the release of the \textit{equivalent heat}.\\

It is of interest to have the above discussion in the reverse order. We note that the entropy can be linked with the volume by the equation: $\Delta S_{X\rightarrow C}=\int dQ/T=-\int_{V_{2}}^{V_{3}}(P/T)dV=-nR\log (V_{3}/V_{2})$. In the classic-thermodynamic point of view, $V$ is a measure of the total number of cells (hence the number of states) which can be occupied by the molecules. There is a basis of the Boltzmann formula, $S=k\log W$, in the Carnot cycle.


\end{document}